\documentclass[11pt,twoside,english]{article}
\usepackage[T1]{fontenc}
\usepackage[latin1]{inputenc}
\usepackage{geometry}
\usepackage{array}
\usepackage{graphicx}
\usepackage{amssymb}

\makeatletter

\newcommand{\noun}[1]{\textsc{#1}}
\providecommand{\tabularnewline}{\\}

\newcommand{\lyxaddress}[1]{
\par {\raggedright #1
\vspace{1.4em}
\noindent\par}
}

\date{}

\usepackage{babel}
\makeatother
\begin{document}
\vspace{4cm}

\title{S\noun{EARCH} FOR THE $\theta_{13}$ NEUTRINO MIXING ANGLE USING
REACTOR ANTI-NEUTRINOS}

\author{D. MOTTA}

\maketitle

\lyxaddress{\begin{center}CEA/Saclay, DSM/DAPNIA/SPP, 91191 Gif-sur-Yvette, France\par\end{center}}

\begin{abstract}
The measurement of the last undetermined neutrino mixing angle $\theta_{13}$
is the main goal of the future experimental research on neutrino oscillations.
At present, $\theta_{13}$ is only known to be much smaller than the
two other mixing angles, $\theta_{12}$ and $\theta_{23}$. The present
bound, which is dominated by the result of the CHOOZ reactor experiment,
is $\sin^{2}2\theta_{13}<0.08$ (at $90\,\%$ confidence level). However,
it is widely recognized that the potential of reactor anti-neutrino
disappearance experiments has not been fully exploited yet. A rich
experimental program is underway, which aims at exploring in the near
future up to $\sin^{2}2\theta_{13}\lesssim0.01$. The targeted sensitivity
requires a clear-cut strategy to reduce significantly both statistical
and systematical errors with respect to past reactor experiments.
A key feature for the success of all projects is the installation
of one or more near identical detectors. The experimental concept
and the status of the upcoming or proposed reactor experiments, and
as well the prospects of the reactor-based search for $\theta_{13}$
are reviewed.
\end{abstract}

\section{Introduction}

In the last decade our understanding of the fundamental properties
of neutrinos has made formidable progress. A variety of experiments
measuring solar, atmospheric, reactor and accelerator neutrinos have
indisputably shown that neutrinos produced in a specific flavor \emph{eigenstate}
oscillate to other flavors. This is proof that neutrinos have mass
and that the weak interaction mixes the mass eigenstates. Experiments
with sensitivities spanning several orders of magnitudes in the oscillation
parameter space have pinned down the values of two mass-squared differences
and two mixing angles ($\theta_{12}$ and $\theta_{23}$). For a review
of the current status of the field, see \cite{Fogli}. On the third
mixing angle $\theta_{13}$, experiments have only set an upper bound.
This parameter translates the admixtures of $\nu_{3}$ in $\nu_{e}$
and hence drives the amplitude of the still unobserved $\nu_{x}\rightarrow\nu_{y}$
oscillations (where \emph{x} or \emph{y} is \emph{e}) at the oscillation
length determined by $\Delta\mbox{m}_{13}^{2}$. At present, $\theta_{13}$
is only known to be much smaller than the two other mixing angles.
The negative result of the Chooz reactor experiment \cite{Chooz}
has provided the most stringent upper-bound %
\footnote{A less stringent bound has been given by the other $\sim1\,\textrm{km}$
baseline reactor experiment, Palo Verde \cite{Palo Verde}%
}: \begin{equation}
\sin^{2}2\theta_{13}\lesssim0.12-0.20\;(90\,\%\:\mbox{C.L.})\label{eq:chooz_limit}\end{equation}
where the interval corresponds to the {}``atmospheric'' mass splitting
allowed by the combined results of Super-Kamiokande, K2K and the first
results of MINOS: $\Delta\mbox{m}_{13}^{2}=2.1-3.0\cdot10^{-3}\,\textrm{eV}^{2}$
($2\sigma$ interval) \cite{Chooz,Schwetz}. A global 1-degree-of-freedom
analysis including all available oscillation data, where $\Delta\mbox{m}_{13}^{2}$
and all other oscillation parameters are marginalized away, gives
\cite{Schwetz}: \begin{equation}
\sin^{2}2\theta_{13}<0.08\;(90\,\%\:\mbox{C.L.})\label{eq:best_present_limit}\end{equation}

The $\theta_{13}$ mixing angle - of fundamental theoretical interest
\emph{per se} - turns out to be the key parameter to access experimentally
CP violation in the leptonic sector, which in turn could explain the
matter/anti-matter asymmetry in the universe. Several upcoming experiments
aim at exploring with two different techniques and complementary approaches
a large part of the region currently allowed by CHOOZ. Superbeam experiments
will exploit the appearance channel $\nu_{\mu}\rightarrow\nu_{e}$
by using $\lesssim1\,\textrm{GeV}$ neutrinos and long baselines ($10^{2}-10^{3}\,\mbox{km}$).
The next generation of reactor experiments aim at measuring the disappearance
of $\overline{\nu}_{e}$ of a few MeV on a $\sim1-2\,\textrm{km}$
baseline, with improved sensitivity with respect to CHOOZ. These two
experimental approaches are nicely complementary \cite{Huber06},
as illustrated below.

In the $\nu_{\mu}\rightarrow\nu_{e}$ search, the contribution of
$\theta_{13}$ to the appearance signal is entangled with unknown
parameters, \emph{i.e.} the CP-violating phase and the mass hierarchy
(the sign of $\Delta\mbox{m}_{13}^{2}$). As a consequence, the sensitivity
to $\theta_{13}$ is limited by the correlations between these parameters
in the appearance probability, and by the degeneracies of the solutions.
This would notably degrade the sensitivity in case no oscillation
is observed. On the other hand, for a favorable combination of the
nuance parameters, superbeam experiments have a chance to observe
a signal even for $\sin^{2}2\theta_{13}<0.01$ \cite{Huber03}, which
is out of reach for the upcoming reactor experiments.

Reactor experiments are sensitive to the disappearance $1-P\left(\overline{\nu}_{e}\rightarrow\overline{\nu}_{e}\right)$,
where $P\left(\overline{\nu}_{e}\rightarrow\overline{\nu}_{e}\right)$
is the $\overline{\nu}_{e}$ survival probability after oscillation.
It is shown \cite{Huber03} that to a very good approximation the
sought signal is given by:\begin{equation}
1-P\left(\overline{\nu}_{e}\rightarrow\overline{\nu}_{e}\right)\simeq\sin^{2}2\theta_{13}\sin^{2}\frac{\Delta\mbox{m}_{13}^{2}\mbox{L}}{4\mbox{E}}\label{eq:disapp}\end{equation}
where L is the oscillation baseline and E the anti-neutrino energy.
Eq. \ref{eq:disapp} shows that for reactor experiments the only unknown
parameter controlling the signal is $\theta_{13}$. This means that
for a source-detector distance tuned to have the oscillatory term
near the maximum ($\mbox{L}\sim1-2\,\mbox{km}$), the disappearance
provides the most clean measurement of $\theta_{13}$. For a small
mixing angle, however, the disappearance of a tiny fraction of anti-neutrinos
can be easily hidden by statistical fluctuations and especially by
the systematic uncertainties on the number of expected events without
oscillation. The reduction of the latter systematics is therefore
the crucial challenge of all future reactor experiments.

\section{Reactor-based search for $\theta_{13}$}

\subsection{The past benchmark: CHOOZ}

\begin{figure}
\begin{centering}\includegraphics[width=0.9\columnwidth]{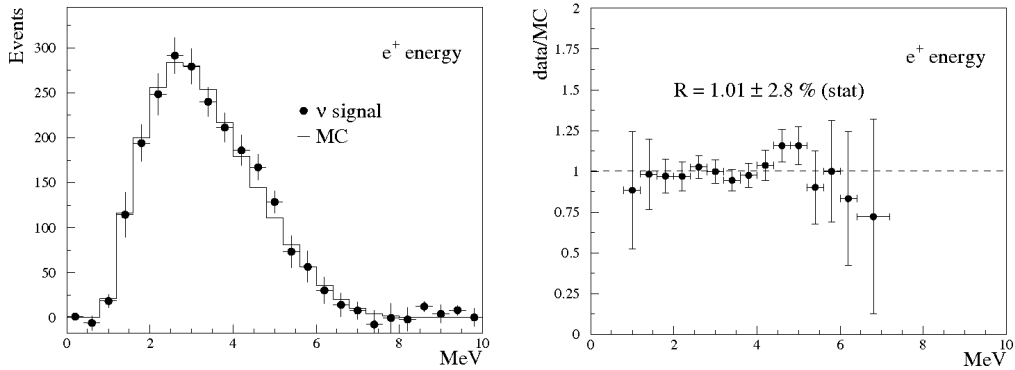}\par\end{centering}

\caption{{\small Left: Signal measured in CHOOZ \protect\cite{Chooz}, after
background subtraction (markers with error bars), superimposed on
the expected spectrum for the case of no oscillation. Right: ratio
of data to predictions.\label{fig:Chooz_results}}}
\end{figure}

CHOOZ \cite{Chooz} has been the last of a long series of short-to-middle
baseline oscillation experiments run in the vicinity of a reactor
power plant. The Chooz power station is located in the Ardennes region,
northeast of France, very close to the Belgian border. The power plant
consists of the two most powerful pressurized water reactors in the
world, capable of yielding up to a total $8.5\,\mbox{GW}$ of thermal
power %
\footnote{The experiment took data during the start-up phase of the nuclear
plant, hence a reduced power was available.%
}. A single detector with a $\sim5\,\mbox{t}$ liquid scintillator
target was located at an average distance from the two cores of $1.05\,\mbox{km}$.
The experiment was stopped after a few months of data-taking, resulting
in a total statistics of $\sim2700\,\nu$ events. The measurement
was fully consistent with the expectations for the case of no oscillation
(Fig. \ref{fig:Chooz_results}). %
 The result can be summarized by the integrated rate normalized to
the MC prediction:\begin{equation}
\mbox{R}=1.01\pm2.8\mbox{\,\%\,(stat)}\pm2.7\,\mbox{\%\,(sys)}\label{eq:Chooz_R}\end{equation}
In addition, Fig. \ref{fig:Chooz_results} shows no evidence of a
spectral deformation, which would be another strong signature of the
oscillation (see Eq. \ref{eq:disapp}). This result translates to
the limit of Eq. \ref{eq:chooz_limit} when combined with the present
knowledge of $\Delta\mbox{m}_{13}^{2}$, and provides the dominant
information for the best limit obtained with a global 2-d.o.f. fit,
Eq. \ref{eq:best_present_limit}.

\subsection{The strategy for the next generation reactor experiments}

The sensitivity achieved by CHOOZ on a possible suppression of the
integrated $\overline{\nu}_{e}$ flux is of the order of $\sim5\,\%$,
as shown by Eq. \ref{eq:Chooz_R}. In the past years a large consensus
has grown on the idea that the potential of a reactor-based search
for $\theta_{13}$ is far from fully exploited \cite{Martemianov,White_paper}.
The present limit suffers from the large statistical and systematical
uncertainties in CHOOZ, which was an experiment conceived for probing
large mixing.

First of all, the next generation experiments need to improve statistics.
This requires an optimal combination of high reactor thermal power,
large target mass, good efficiency and long run-time. Though trivial
in principle, the latter requirements are not {}``painless'': projects
envisaging very large target masses are expensive; a long data-taking
($\sim3\,\mbox{y}$) demands an extremely stable and well monitored
detector.

However, most challenging is the reduction of the systematic uncertainties:
while it seems realistic to reduce the statistical error to $<0.5\,\%$,
the control of systematics to this level is not possible with an approach
{}``à la CHOOZ''. Most importantly, because the ultimate knowledge
of the neutrino source (the reactors) is limited to $\sim2\,\%$.
Secondly, because several detector parameters influencing the neutrino-rate
can be determined with a poor accuracy even after calibrations, like
the number of target protons, efficiency of the analysis cuts, capture
on Gd versus H, $\nu$-produced neutrons spilling outside or inside
the target volume, etc. A more detailed discussion about these and
other systematics follows in Sec. \ref{sub:Systematics}. 

For the above reasons, all of the next generation projects propose
to deploy one or several near detectors, strictly identical to the
detector that searches for the oscillation. These near detectors will
monitor the nearly-unoscillated flux and spectrum with {}``identical''
response to that of the far detector. Most of the reactor-related
uncertainties are correlated between detectors and hence perfectly
cancel, and many the detector systematics are reduced as well, provided
that the response of near and far detectors are identical (after calibration).
An additional, challenging option proposed, for example, for the Daya
Bay experiment, consists in building several movable detectors, which
would be periodically swapped between near and far sites. Provided
that this operation can be done reliably and without affecting the
detector response, some of the remaining systematics would cancel
in this way. The sensitivity of the next generation experiments will
ultimately depend on the accuracy within which two detectors can be
made identical, or to which extent their relative response can be
known.

\subsection{Experimental concept}

As in past reactor experiments, $\overline{\nu}_{e}$ are detected
via the reaction:\begin{equation}
\overline{\nu}_{e}+p\rightarrow e^{+}+n\label{eq:nu+p}\end{equation}
Eq. \ref{eq:nu+p} has a threshold of $1.8\,\textrm{MeV}$. The neutron
carries a very little kinetic energy, hence the positron energy deposition,
boosted by the $1.024\,\textrm{MeV}$ from its annihilation, measures
the $\overline{\nu}_{e}$ energy. An H-rich organic liquid scintillator
is used as target and detection medium. The neutron issued from the
interaction thermalizes in the scintillator and is then captured.
The resulting delayed gamma emission provides a powerful signature
to reject the background. 

In order to enhance the background rejection, Gd is dissolved in the
scintillator at a concentration of $\sim1\,\textrm{g/l}$. Captures
on $^{155}\textrm{Gd}$ and $^{157}\textrm{Gd}$, which have huge
cross sections for thermal neutrons, are followed by the emission
of a $\sim8\,\textrm{MeV}$ gamma cascade, well beyond the background
due to the natural radioactivity. A typical $\overline{\nu}_{e}$
event is then given by the coincidence of the prompt $e^{+}$ and
a delayed $\sim8\,\textrm{MeV}$ energy deposition, within a time
window of $\sim100\,\mu s$ (the characteristic n capture time for
$\sim1\,\textrm{g/l}$ Gd concentration is $\sim30\,\mu s$).

\begin{figure}
\begin{centering}\includegraphics[width=10cm,height=6.5cm]{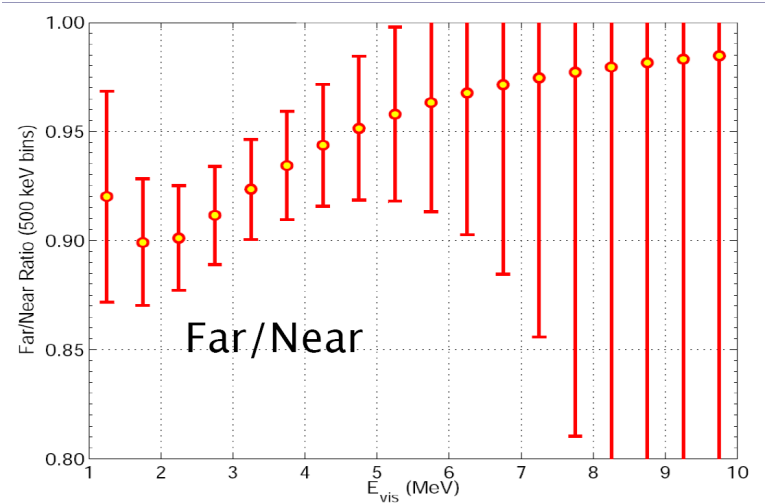}\par\end{centering}

\caption{{\small Far/Near rate ratio in Double Chooz after 3y data-taking,
for $\sin^{2}2\theta_{13}=0.1\,,\,{\Delta m}_{13}^{2}=2.5\cdot10^{-3}\,\textrm{eV}^{2}$.
Error bars are statistical. \label{fig:DC_signal}}}
\end{figure}

A non vanishing value of $\theta_{13}$ would show up in the data
as an energy-dependent suppression of the rate in the far detector
with respect to the near detector. It can be convenient to separate
the experimental information giving sensitivity to $\theta_{13}$
in two contributions: an integrated spectrum-averaged suppression
and a spectral deformation. It will be later shown that the relative
weight of the two can vary, depending on the characteristics of the
experiment. As an example, Fig. \ref{fig:DC_signal} shows the kind
of oscillation signal that would be seen in Double Chooz \cite{DC_proposal}
(described in Sec. \ref{sub:Double-Chooz}) for a relatively large
$\theta_{13}$ ($\sin^{2}2\theta_{13}=0.1$) and $\Delta\mbox{m}_{13}=2.5\cdot10^{-3}\,\textrm{eV}^{2}$.%

\begin{figure}
\begin{centering}\includegraphics[width=0.5\columnwidth]{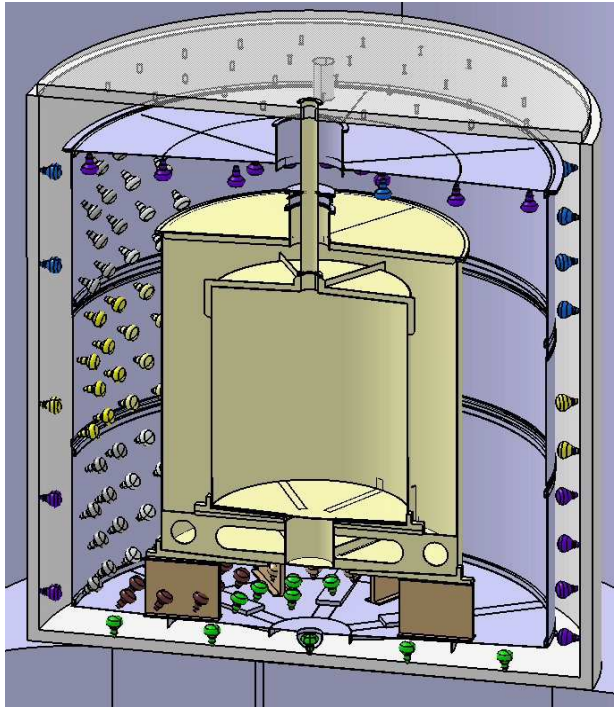}\par\end{centering}

\caption{{\small Sketch of the Double Chooz detector. Other projects propose
very similar designs.\label{fig:DC_detector}}}
\end{figure}

The detector design is an evolution of CHOOZ. Owing to what has been
learned in the past, the main improvements concern the minimization
of the information losses, (dead volumes, non-fully-contained events,
badly reconstructed events) and the reduction of the backgrounds (overburden,
$\mu$-veto-systems, shielding, material radio-purity). As an example,
Fig. \ref{fig:DC_detector} illustrates the design of the Double Chooz
detector (all other projects propose very similar concepts). %
 From the center, the following sub-systems are typically found:

\begin{enumerate}
\item A central cylindrical transparent vessel containing several tons of
liquid scintillator doped with Gd at $\sim1\,\textrm{g/l}$. This
is the target for the $\overline{\nu}_{e}$ interactions.
\item A volume concentric with the target and containing undoped scintillator.
This region, called {}``$\gamma$-catcher'', acts as a calorimeter:
its purpose is to assure the full collection of the positron energy,
and a good containment of the gamma cascade following n-capture on
Gd.
\item A non-scintillating, highly transparent liquid buffer (mineral oil)
contained in a stainless steel tank, on which low-background photomultiplier
tubes (PMT) are mounted to detect the scintillation photons. The buffer
volume is one of the main improvements with respect to CHOOZ. It attenuates
the gammas emitted by the PMTs, and thus reduces the rate of accidentals.
\item An instrumented volume, called $\mu$-veto, to tag and reject events
induced by cosmic muons.
\item An external shielding to protect the innermost volumes from the gammas
of the natural radioactivity in the surrounding rocks.
\end{enumerate}
A system of concentric chimneys will allow the introduction of the
filling tubes and non-permanent calibration devices.

Near and far detector both need to be deployed underground to insure
adequate protection against the cosmic rays, which would produce an
overwhelming background at surface. A discussion about backgrounds
will follow in Sec. \ref{sub:Backgrounds}.

\subsection{Sensitivity\label{sub:Sensitivity}}

In this section some general remarks about the sensitivity of the
reactor-based search for $\theta_{13}$ are presented. The discussion
is based on the work published in \cite{Huber06} and earlier by the
same group.

\begin{figure}
\begin{centering}\includegraphics[width=12cm,height=9cm]{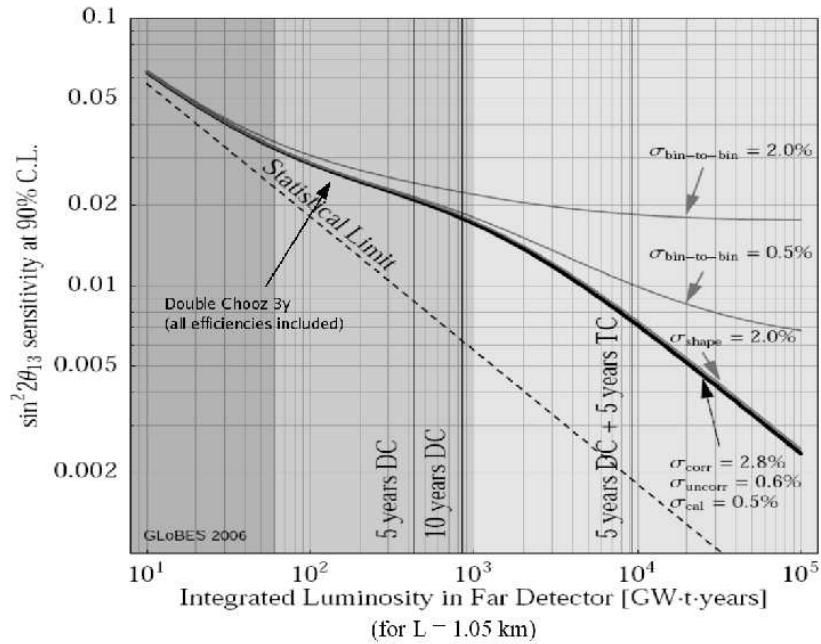}\par\end{centering}

\caption{{\small Sensitivity of a reactor experiment to $\sin^{2}2\theta_{13}$,
as a function of the total cumulative statistics. The original plot
is from \protect\cite{Huber06}. Here the three regimes discussed
in the text are highlighted with different levels of gray. For reference,
the statistics of Double Chooz realistically achievable after 3 years
of data-taking is indicated.\label{fig:sensitivity}}}
\end{figure}

The sensitivity, here defined as the largest value of $\sin^{2}2\theta_{13}$
that would be consistent with data at a certain confidence level (typically
90 \%) if no oscillation is observed, depends on both the statistical
and systematical errors. The latter can be separated in three terms:
an error completely correlated between near and far detector(s) ($\sigma_{corr}$),
the relative global normalization uncertainty between the two detectors
($\sigma_{norm}$), and a bin-to-bin error non-correlated between
energy bins and between detectors ($\sigma_{bin-to-bin}$). Fig. \ref{fig:sensitivity}
shows the evolution of the sensitivity with the total statistics in
the far detector.%
 With all systematic effects switched off, the sensitivity would follow
a trivial $\propto\sqrt{N}$ statistical law. An experiment of the
size of Double Chooz ($8.2\,\mbox{t}$) would then be sensitive to
$\sin^{2}2\theta_{13}\sim0.015$. When the systematics are turned
on, three {}``regimes'' appear: a {}``low statistics regime'',
where the sensitivity is dominated by the rate ratio and is statistics-limited;
an {}``intermediate statistics regime'', where both rate ratio and
spectral deformation contribute to the sensitivity, the former limited
by $\sigma_{norm}$, the latter by the statistics per energy bin;
and a {}``high statistics regime'', where most of the sensitivity
comes from the spectral deformation and is hence limited by $\sigma_{bin-to-bin}$.
The transition between these regimes depends on the size of the systematics.
However, several general conclusions can be drawn.

The sensitivity on any experiment will first rapidly roll along the
curve of Fig. \ref{fig:sensitivity} (note the log-log scale), and
will eventually reach a plateau, which is ultimately set by statistics
or systematics. For a small, {}``first generation'' detector ($\sim10\,\mbox{t}$),
the {}``roll-down'' will stop at the second regime. In this case
the final sensitivity critically depends on $\sigma_{norm}$ and the
major experimental efforts need to be dedicated to the reduction of
this systematics. Fig. \ref{fig:sensitivity} shows that, for a constant
$\sigma_{norm}$, a significant improvement of the sensitivity requires
fully entering the third regime, corresponding to a $\gtrsim200\,\mbox{t}$
target mass. A {}``second generation'' experiment of intermediate
size ($\sim50-100\,\mbox{t}$) will therefore need also an improvement
of $\sigma_{norm}$ to outperform a smaller detector. A very large
detector, instead, could afford ignoring $\sigma_{norm}$ and simply
rely on the analysis of the spectral distortion. In this case the
systematic to keep under control is $\sigma_{bin-to-bin}$, hence
any possible energy-dependent difference between the signals at the
near and far sites (e.g. due to the background subtraction at different
sites, or biased analysis cuts).

\subsection{Systematics\label{sub:Systematics}}

The previous section has outlined how critically the systematical
errors affect the ultimate sensitivity of the reactor-based search
for $\theta_{13}$. Table \ref{tab:systematics} presents a summary
of all identified systematics, as they were measured in CHOOZ, as
they are estimated after the R\&D carried out in Double Chooz, and
as they are foreseen in Daya Bay (Sec. \ref{sub:Daya-Bay}).%
\begin{table}

\caption{{\small Summary of the systematic uncertainties. The {}``realistic
now'' column is based on assessments in the Double Chooz proposal
\protect \cite{DC_proposal} and on the {}``baseline'' uncertainties
estimated in Daya Bay \protect \cite{DayaBay}. The {}``wished for
the future'' column reflects the Daya Bay R\&D goals (without detector
swapping).\label{tab:systematics}}}

\vspace{0.4cm}

\begin{centering}\begin{tabular}{|c|c|c|c|c|}
\hline 
&
&
CHOOZ&
realistic now&
wished for future\tabularnewline
\hline
\hline 
&
Power&
$\sim2\,\%$&
negligible&
negligible\tabularnewline
&
\multicolumn{1}{c|}{E/fission}&
0.6 \%&
negligible&
negligible\tabularnewline
Reactor&
$\overline{\nu}_{e}$/fission&
0.2 \%&
negligible&
negligible\tabularnewline
&
$\sigma$&
0.1 \%&
negligible&
negligible\tabularnewline
&
Distances&
negligible&
0.1 \%&
0.1 \%\tabularnewline
\cline{2-2} \cline{3-3} \cline{4-4} \cline{5-5} 
\multicolumn{1}{|c|}{}&
Tot Reactor&
$\sim2.1\,\%$&
0.1 \%&
0.1 \%\tabularnewline
\hline
\hline 
\# target p&
&
0.8 \% &
0.2 \% &
0.1 \%\tabularnewline
&
&
(absolute)&
(relative)&
(relative)\tabularnewline
\hline
\hline 
&
$\mbox{e}^{+}$ energy cut&
0.8 \%&
0.1 \%&
0.05 \%\tabularnewline
&
Gd/H captures&
1.0 \%&
0.2 \%&
0.1 \%\tabularnewline
&
n energy cut&
0.4 \%&
0.2 \%&
0.1 \%\tabularnewline
Efficiency&
$\mbox{e}^{+}$- n distance&
0.3 \%&
not used&
not used\tabularnewline
&
$\mbox{e}^{+}$- n delay&
0.4 \%&
0.1 \%&
0.03 \%\tabularnewline
&
n multiplicity&
0.5 \%&
negligible&
negligible\tabularnewline
&
dead-time&
negligible&
negligible&
negligible\tabularnewline
\cline{2-2} \cline{3-3} \cline{4-4} \cline{5-5} 
\multicolumn{1}{|c|}{}&
Tot efficiency&
1.5 \%&
0.3\%&
0.15 \%\tabularnewline
\hline
\hline 
&
Grand Total&
2.7 \%&
0.4 \%&
0.2 \%\tabularnewline
\hline
\end{tabular}\par\end{centering}
\end{table}

The use of identical near detectors suppresses the source-related
systematics from 2.1 \% to $0.1\,\%$. This remaining $0.1\,\%$ comes
from the error of the reactors-to-near detector(s) distance, including
the the finite size of reactors and detectors, and the fluctuations
of the $\overline{\nu}_{e}$ production barycenter. Experiments installed
in a multi-core (> 2) site, like Daya Bay and RENO (Sec. \ref{sub:RENO}),
suffer also from the additional uncertainty due to the different repartition
of the signal from each single reactor in the near detector(s), compared
to the far detector.

As regards the number of target protons, in CHOOZ this error was quite
large (0.8 \%). An important improvement is guaranteed by the use
of the same batch of liquid scintillator in the near and far detectors.
This assures that the uncertainty due to the chemical composition
(H mass fraction) is canceled out. The only contribution left comes
from the relative error between mass or volume measurements. Furthermore,
the use of identical near and far detectors cancels the {}``spill-in,
spill-out'' error (imperfect compensation between $\nu$-events in
the $\gamma$-catcher with a n spilling into the target volume, and
$\nu$-events in the target with the n spilling out to the $\gamma$-catcher),
which is equivalent to an additional uncertainty on the target proton
number.

A strong reduction is also envisaged for the detector-related systematics,
i.e. the detection efficiency. In this case as well, the multi-detector
concept transforms uncertain absolute efficiency determinations to
better constrained relative comparison. The use of the same scintillator
and calibration devices in detectors of identical geometry will be
crucial. Moreover, a significant reduction is assured by the improvement
of the detector design, which will allow to limit the number of analysis
cuts. In fact, several {}``quality'' cuts were needed in CHOOZ to
suppress the accidentals (Sec. \ref{sub:Backgrounds}). A simplified
event selection will be possible thanks to a reduction of the rate
of singles (better shielding) and with an improved containment of
the $\mbox{e}^{+}$ events and Gd cascades.

In conclusion, the reduction of the total systematic error from 2.7
\% to $<0.6\,\%$ (the Double Chooz conservative goal) seems realistic
nowadays. The goal of $<0.2\,\%$, required for a significant boost
of the sensitivity in next-generation projects, is not yet validated
by a specific R\&D. At this regard, the upcoming first generation
experiments will set important benchmarks and pave the ground for
a possible further step in sensitivity.

\subsection{Backgrounds\label{sub:Backgrounds}}

Background subtraction is another source of systematics, as it can
both affect the global normalization between near and far detectors
and as well the spectral shape. CHOOZ had the chance to measure the
background \emph{in situ} with both reactors off, as it started data-taking
just before the power plant commissioning. The chance for a {}``signal-free''
background measurement depends on the characteristics of the site.
In a 2-core plant, like in Double Chooz, a long simultaneous stop
of both reactors is extremely rare, but not impossible on a 3y time
scale. For a multi-core site, this event is nearly impossible and
one can simply expect a modulation of the total power %
\footnote{It must be noted that single reactors are periodically off once a
year during $\sim$2 weeks for refueling.%
}.

Three different classes of background can be identified: accidentals,
due to the random coincidence of an environmental gamma and a n-like
event within the $\nu$ time window; fast neutrons, which can produce
proton-recoils in the target (misidentified as $\mbox{e}^{+}$), and
then be captured; and long-lived cosmogenic radio-isotopes with significant
branching for $\beta$ -decay followed by neutron emission ($^{9}\mbox{Li}$,
$^{8}\mbox{He}$). 

Accidentals can be measured \emph{in situ} even with reactors on,
through the rates and spectra of the singles. In order to reduce the
subtraction error, the rate of gammas penetrating the target must
be reduced with an accurate material selection (low radioactive contaminants)
and an adequate shielding. 

Fast neutrons produce a background with very similar signature as
the $\nu$-events, however with a nearly flat spectrum. Hence the
rate can be extrapolated from the part of the spectrum at higher energy
than the signal. Since fast neutrons are secondary particles produced
by cosmic muons, a large and effective $\mu$-veto is very helpful.

Stopping and showering $\mu$ can originate spallation reactions on
$^{12}\mbox{C}$ that produce $^{9}\mbox{Li}$ and $^{8}\mbox{He}$.
These isotopes can decay with a $\beta$-n sequence that is indistinguishable
from a $\nu$ event. Lifetimes are of the order of $\sim10^{2}\,\mbox{ms}$,
which makes it hard to impose a long enough veto to all crossing muons.
A possibility would be developing a more sophisticated $\mu$ detector,
capable of tagging muons as {}``showering'' or {}``stopping''.
A longer veto could be applied to these events only, without increasing
the dead-time too much.

Generally speaking, all the discussed backgrounds rapidly attenuate
with the laboratory overburden. A compromise must be reached between
costs and allowed background rate, bearing in mind that the better
the knowledge of the background, the higher the rate that can be tolerated.
As a rule of thumb, a first generation experiment needs to keep the
background uncertainty per energy bin below $\sim1\,\%$ of the signal.
A second generation experiment should make roughly two times better.

\section{Upcoming and proposed experiments}

By the time these proceedings are written, only three experiments
have survived the decisions of their funding agencies: Double Chooz,
Daya Bay and RENO. After Braidwood was turned down by the negative
decision of the DOE in April 2006, another project, the Japanese KASKA,
has been stopped upon rejection of its funding requests. The KASKA
collaboration has since then joined Double Chooz.

\subsection{Double Chooz\label{sub:Double-Chooz}}

Double Chooz \cite{Dario_DC_Vietnam,DC_proposal} is the most advanced
among these projects. The R\&D is essentially completed, the detector
design finalized and the project is now fully entering the construction
phase. Double Chooz will be installed nearby the same nuclear power
plant of the CHOOZ experiment. Table \ref{tab:DC_synopsis} shows
a synopsis of the experiment.%
\begin{table}

\caption{{\small Double Chooz synopsis. The sensitivity in the first row refers
to the first experimental phase, in which the far detector will run
alone.\label{tab:DC_synopsis}}}

\vspace{0.4cm}

\begin{centering}\begin{tabular}{|>{\centering}p{0.09\columnwidth}||>{\centering}m{0.1\columnwidth}|>{\centering}m{0.14\columnwidth}|>{\centering}m{0.13\columnwidth}|>{\centering}m{0.08\columnwidth}|>{\centering}m{0.06\columnwidth}|>{\centering}m{0.05\columnwidth}|>{\centering}m{0.11\columnwidth}|}
\hline 
Reactors&
Detectors {\scriptsize ($\mbox{M}=8.2\,\mbox{t}$)}&
Sites&
Civil works&
Stat {\small ($\mbox{d}^{-1}$) (incl. $\epsilon$)}&
B/S {\small (\%)}&
Start&
Sensitivity ($\sin^{2}2\theta_{13}$)\tabularnewline
\hline 
\multicolumn{1}{|p{0.09\columnwidth}||}{\parbox[t]{0.09\columnwidth}{1 site\\
2 cores}}&
1 Far&
$\mbox{D}=1.05\,\mbox{km}$ ($300\,\mbox{mwe}$)&
refurbishing&
$50$&
$\lesssim1$&
2008&
$0.06$

(1.5 y)\tabularnewline
\cline{2-2} \cline{3-3} \cline{4-4} \cline{5-5} \cline{6-6} \cline{7-7} \cline{8-8} 
\multicolumn{1}{|p{0.09\columnwidth}||}{$8.5\,\mbox{GW}_{th}$ }&
1 Near&
$\mbox{D}\sim280\,\mbox{m}$ ($\sim80\,\mbox{mwe}$)&
$\sim40\,\mbox{m}$ shaft + lab&
$550$&
$\lesssim0.5$&
2010&
$<0.03$

(3 y)\tabularnewline
\hline
\end{tabular}\par\end{centering}
\end{table}

One of the main advantages of Double Chooz is the availability of
the CHOOZ laboratory for the integration of the far detector, which
allows the experiment to be faster and cheaper than any other project.
As regards the near site, the present baseline in the discussions
with EDF assumes the construction of a laboratory at $\sim280\,\textrm{m}$
from the cores, with an overburden of $\sim80\,\textrm{mwe}$.

The Double Chooz collaboration is composed of several institutions
from Europe (France, Germany, England, Russia and Spain), US and Japan.
The project is fully approved and funded in Europe, awaits approval
in US and Japan in 2007. A close collaboration with the power plant
management has been established, both for the refurbishing and recommissioning
of the far laboratory, and the construction of the near laboratory.
The current planning considers material procurement to be completed
by 2007, with the start of the integration of the far detector in
the fall of 2007 and commissioning at the beginning of 2008. Parallel
to this phase, the construction of the near laboratory will proceed.
The site is expected to be available for detector integration in 2009.
Double Chooz will then be operative with both detectors by 2010.

The intense R\&D work carried out by the Double Chooz collaboration
has validated the robustness of the detector concept and the feasibility
of the sensitivity goals. In the first phase with the far detector
alone, Double Chooz will attain a sensitivity to $\sin^{2}2\theta_{13}$
of $0.06$, and then will reach its goal sensitivity of better than
$0.03$ by 2013.

\subsection{Daya Bay\label{sub:Daya-Bay}}

Daya Bay is a project aiming at a final sensitivity of $\sim0.01$.
It is based on a strong US-Chinese collaboration that also includes
Russian, Taiwanese and Czech institutions. It is approved in China
and strongly supported by the DOE in US. The experiment will be installed
nearby the Daya Bay / Ling Ao nuclear power plants, about 55 km away
from Hong Kong.

In order to achieve the desired sensitivity, Daya Bay relies on a
number of ambitious design options and merit factors of its site.
The mass scale of the experiment will be of one order of magnitude
higher than Double Chooz, with a total target mass of 160 t of Gd-loaded
scintillator distributed among one far and two near sites. The Daya
Bay site has ideal geology, with nearby mountains for very deep underground
laboratories. The experiment will benefit from an intense total $\overline{\nu}_{e}$
flux, however at the cost of a non trivial reactor arrangement. Four
reactors are distributed on 2 separate sites $1\,\mbox{km}$ away
from each other, and a third 2-reactor site will be added by 2011.

The proposed design envisions splitting the total target mass in $8\times20\,\mbox{t}$
modules (each weighting $\sim90\,\mbox{t}$ with $\gamma$-catcher
and buffer) placed side-by-side in big water pools measuring $16\times16\times10\,\mbox{m}^{3}$,
used as both shielding and water-Cherenkov veto; 4 modules in the
far laboratory, and 2 in each of the two near detectors. An option
the collaboration bets on is the possibility to make all modules movable
on rails between experimental sites. The synopsis of the Daya Bay
project is shown in Table \ref{tab:DB_synopsis}. %
\begin{table}

\caption{{\small Daya Bay synopsis. Statistics given with the 3 sites and
6 reactors on, for near and far sites, 2/4 for the mid site ($11.6\,\mbox{GW}_{th}$).
Note that a similar sensitivity as Double Chooz is claimed for the
Mid detector, even though the contribution of the Ling Ao I reactors,
which will account for more than half of the signal, will not be monitored
by the corresponding near detector. \label{tab:DB_synopsis}}}

\begin{centering}\vspace{0.4cm}
\begin{tabular}{|>{\centering}m{0.1\columnwidth}||>{\centering}m{0.1\columnwidth}|>{\centering}m{0.15\columnwidth}|>{\centering}m{0.13\columnwidth}|>{\centering}m{0.06\columnwidth}|>{\centering}m{0.06\columnwidth}|>{\centering}m{0.05\columnwidth}|>{\centering}m{0.11\columnwidth}|}
\hline 
Reactors&
Detectors {\footnotesize ($\mbox{M}=20\,\mbox{t}$)}&
Sites&
Civil works&
Stat {\small ($\mbox{d}^{-1}$)}&
B/S {\small (\%)}&
Start&
Sensitivity ($\sin^{2}2\theta_{13}$)\tabularnewline
\hline
\hline 
\multicolumn{1}{|p{0.1\columnwidth}||}{\parbox[t]{0.1\columnwidth}{3 sites\\
6 cores\\
}}&
\parbox[c]{0.1\columnwidth}{4 Far\\
(1 site)}&
$\mbox{D}\sim1.8\,\mbox{km}$ ($\sim1000\,\mbox{mwe}$)&
$\sim1\,\mbox{km}$ tunnel + 1 cavity&
$200$&
$\lesssim0.5$&
2010&
$0.01$

(3 y)\tabularnewline
\cline{2-2} \cline{3-3} \cline{4-4} \cline{5-5} \cline{6-6} \cline{7-7} \cline{8-8} 
\multicolumn{1}{|p{0.1\columnwidth}||}{}&
\parbox[c]{0.1\columnwidth}{2 Mid\\
(1 site)}&
$\mbox{D}\sim1\,\mbox{km}$ ($\sim550\,\mbox{mwe}$)&
$\sim1\,\mbox{km}$ tunnel + 1 cavity&
$200$&
&
2009&
$0.035$

(1 y)\tabularnewline
\cline{2-2} \cline{3-3} \cline{4-4} \cline{5-5} \cline{6-6} \cline{7-7} \cline{8-8} 
\multicolumn{1}{|p{0.1\columnwidth}||}{$17.4\,\mbox{GW}_{th}$ }&
\parbox[c]{0.1\columnwidth}{4 Near\\
(2 sites)}&
$\mbox{D}\sim400\,\mbox{m}$ ($\sim300\,\mbox{mwe}$)&
$\sim1\,\mbox{km}$ tunnels + 2 cavities&
$2000$&
$\lesssim0.5$&
&
\tabularnewline
\hline
\end{tabular}\par\end{centering}
\end{table}

Daya Bay is now at a R\&D and preparation stage. This phase will have
to finalize the detector design and answer several questions concerning
a realistic estimation of the systematical errors and the feasibility
of the swapping option. For its ambitious goals and mass scale, Daya
Bay should be considered a {}``second generation'' project. The
collaboration, however, is trying hard to be competitive as a first
generation experiment as well, by deploying two additional modules
at an intermediate site at $\sim1\,\mbox{km}$ distance with only
one near site on. The mid experiment is supposed to be running while
the rest of the civil works and installations are under completion.
Whether this {}``fast Daya Bay'' option is worth the effort will
be answered by ongoing studies.

\subsection{RENO\label{sub:RENO}}

RENO\cite{RENO} is a $20\,\mbox{t}$ scale experiment proposed by
a South-Korean collaboration, to be installed at the Yonggwang nuclear
power plant. The site has favorable geology with nearby hills potentially
providing adequate overburden for near and far detector. The total
$\overline{\nu}_{e}$ flux is 3 times higher than in Chooz, however
this flux is produced by 6 reactors lined up over $1.5\,\mbox{km}$.
As a consequence, a large part of the reactor-related systematics
cannot be canceled by the near detector (see Table \ref{tab:RENO_reactor_contributions}).%
\begin{table}

\caption{{\small Relative contribution (\%) of the 6 reactors in the proposed
RENO experiment.\label{tab:RENO_reactor_contributions}}}

\begin{centering}\vspace{0.4cm}
\begin{tabular}{|c||c|c|c|c|c|c|}
\hline 
&
R1&
R2&
R3&
R4&
R5&
R6\tabularnewline
\hline 
Near&
3.0&
7.8&
39.2&
39.2&
7.8&
3.0\tabularnewline
\hline 
Far&
14.8&
16.9&
18.3&
18.3&
16.9&
14.8\tabularnewline
\hline
\end{tabular}\par\end{centering}
\end{table}
 The RENO collaboration proposes therefore to install smaller {}``very
near'' detectors to monitor the neutrino flux of each reactor. The
claimed sensitivity would then be $\sim0.02$.

RENO is still at a very early stage of the R\&D, however the Korean
funding agencies seem very supportive.

\subsection{Ideas for the future}

Recalling the discussion in Sec. \ref{sub:Sensitivity}, the sensitivity
of the upcoming experiments Double Chooz and Daya Bay will be ultimately
limited by the error of the relative normalization between near and
far detector(s). Decreasing this systematics further with respect
to the very challenging goal of $\sigma_{rel}<0.2\%$ does not seem
realistic for the moment. To push the sensitivity beyond the barrier
of $\sin^{2}2\theta_{13}<0.01$ it is necessary to increase the statistics
so much to enter the third regime of Fig. \ref{fig:sensitivity},
where the spectral shape distortions dominate the sensitivity. Provided
that the measurement of the $\overline{\nu}_{e}$ spectra can be made
with the required accuracy and that after Double Chooz, Daya Bay and
the superbeam experiments the question about $\theta_{13}$ is still
of the highest scientific priority, the only viable strategy seems
installing a $\gtrsim200\,\mbox{t}$ detector at the optimal distance
from a reactor site, shielded by a very good overburden. The near
detector is no longer required to be identical, provided that no energy-dependent
biases are introduced in this way. The ANGRA \cite{ANGRA} project
aims at exploring this solution during the forthcoming years. The
detector would be installed by the nuclear power plant of Angra dos
Reis, in Brazil. At present the project is in the phase of conceptual
design and sensitivity study.

\section{Conclusions and prospects}

Double Chooz is fully approved in Europe and has now entered the material
procurement phase. It will be the first experiment to improve the
present limit on $\sin^{2}2\theta_{13}$ and will reach its goal sensitivity
by 2013, roughly when T2K is expected to start releasing competitive
results on $\theta_{13}$. Daya Bay, supported by a very strong US-Chinese
commitment, aims at surpassing the Double Chooz sensitivity with a
bigger and more precise experiment. The realistically attainable systematics,
the related sensitivity, the feasibility of the proposed challenging
solutions, and the cost and schedule of the project are currently
under investigation. 

Most theoretical models are unable to explain why $\theta_{13}$ should
be much smaller than the two other mixing angles. If the small (1,3)
$\nu$-mixing is merely accidental, $\theta_{13}$ is most likely
not very far from the present bound and will be measured within few
years from now by Double Chooz, and then confirmed with higher significance
by Daya Bay. Early indications about the CP phase and mass hierarchy
could then be deduced through the comparison with the results of the
superbeams.

A much smaller (1,3) mixing would then be a strong hint for a new
discrete flavor symmetry. The synergies between superbeams and second
generation reactor experiments (class $\sin^{2}2\theta_{13}<0.01$)
will then be very powerful towards the interpretation of the nature.

\end{document}